\documentstyle[12pt]{article}
\newcommand{\be}{\begin{equation}}
\newcommand{\qee}{\end{equation}}
\begin{document}
\begin{titlepage}
\title{
{\bf Loop transfer matrix \\
and\\
gonihedric loop diffusion\\
}
}
{\bf
\author{
T.Jonsson\footnote{Permanent address: University of Iceland, Dunhaga 3, 
107 Reykjavik, Iceland}~~  and ~~G.K. Savvidy\\
National Research Center Demokritos,\\
Ag. Paraskevi, GR-15310 Athens, Greece \\
}
}
\date{}
\maketitle
\begin{abstract}

We study a class of statistical systems which simulate 3D gonihedric system 
on euclidean lattice. We have found the exact partition function of the 3D-model 
and the corresponding critical indices analysing the transfer matrix 
$K(P_{i},P_{f})$ which describes the propagation of loops on a lattice. 
The connection between 3D gonihedric system and 2D-Ising model is clearly seen.

\end{abstract}
\thispagestyle{empty}
\end{titlepage}
\pagestyle{empty}
\pagenumbering{}
\vspace{.5cm}

\section{Introduction}

In the articles \cite{amb} the authors formulated 
a model of random surfaces with an action which is proportional
to the linear size of the surface. The model has a number of properties which 
make it very close to the Feynman path integral for a point-like 
relativistic particle. In the limit when the surface degenerates into a 
single world line, the action becomes proportional to the length of the 
path and the classical equation of motion for the gonihedric string
is reduced to the classical equation of motion for a free relativistic 
particle.\footnote{The problems of spiky instability and the convergence 
of the partition function have been studied in \cite{amb,durhuus,schneider}} 

In addition to the formulation of the theory in the continuum space 
the system allows an equivalent representation on Euclidean lattices 
where a surface is associated with a collection of plaquettes 
\cite{weg,sav}. In these lattice spin systems 
the interface energy coinsides with the linear-gonihedric action for 
random surfaces. This gives an opportunity for analytical investigations 
\cite{sav2,sav1,pie,wegpie,karowski,cappi} and numerical simulations 
\cite{baillie,bathas,johnson,cut,cappi} of the corresponding statistical systems. 

Additional understanding of the physical 
behaviour of the system comes 
from the analysis of the transfer matrix \cite{sav2} which describes the 
propagation of the closed loops-strings in time direction with an 
amplitude which is proportional to the sum of the length of the string  
and of the total curvature. In this article we shall study
the physical picture of string propagation which was 
suggested in the transfer matrix approach \cite{sav2}. 

The partition function of the system is defined  as \cite{amb}
\be
Z_{gonihedric}(\beta) = \sum_{\{M\}} exp\{-\beta A(M) \} ,~~~~~~A(M) = 
\sum_{<ij>} \lambda_{ij}
\cdot \vert \pi - \alpha_{ij} \vert , \label{part}
\qee
where $\lambda_{ij}=a$ is the length of the edge $<ij>$ which is equal 
to the lattice spacing $a$ on the cubic lattice, $\alpha_{ij}$ is the 
dihedral angle 
between two neighbouring plaquettes of the singular surface $M$ 
sharing a common edge $<ij>$, $\alpha_{ij}=0,\pi/2 ,\pi$. 
\footnote{Usually we take
$a=1$.} In (\ref{part}) $\{M\}$ denote the set of closed singular 
surfaces on the three-dimensional toroidal
lattice $T^{3}$ of size $N \times N \times N$. {\it A singular 
surface $M$ is a collection of plaquettes in $T^{3}$ such that every 
link is contained in 0, 2 or 4 plaquettes and every plaquette of the 
lattice $T^{3}$ can be occupied only once}~\cite{weg,sav}. 
The surfaces are closed because only an even
number of plaquettes meet at a given lattice link. The singular surfaces
of interface which describe the states of arbitrary three-dimensional 
spin system can be viewed as a set of surfaces $\{M\}$ \cite{weg,wegner}.

In  \cite{sav2} it has been proven that the partition function 
(\ref{part}) can be represented in the form
\be
Z(\beta) = 
\sum_{\{P_{1},P_{2},...,P_{N}\}}~ K_{\beta}(P_{1},P_{2})\cdots 
K_{\beta}(P_{N},P_{1})  = tr K^{N}_{\beta}\label{trace},
\qee
where $K_{\beta}(P_{1},P_{2})$ is the transfer matrix of  size 
$\gamma \times \gamma $, defined as (see formulas (15),(16) of \cite{sav2}) 
\be
K^{gonihedric}_{\beta}(P_{1},P_{2}) = 
exp \{-\beta ~[k(P_{1}) + 
2 l(P_{1} \bigtriangleup  P_{2}) + k(P_{2})]~ \},  \label{tran}
\qee
where $P_{1}$ and $P_{2}$ are closed polygons on a two-dimensional toroidal 
lattice 
$T^{2}$ of size $N \times N$ and $\gamma$ is the total number of polygon-loops 
on a toroidal lattice $T^{2}$. {\it Closed polygons $\{P\} \equiv \Pi$ 
are associated with the 
collection of links on $T^{2}$ with the restriction that only an even number 
of links can intersect at a given vertex of the lattice and that the links 
can be occupied only once}. \footnote{The polygon-loops $P$ appear as the 
intersection of the 
singular surfaces $M$ with the planes between coordinate planes in $T^3$ 
\cite{sav2}.}

The transfer martix (\ref{tran}) can be viewed as describing  the propagation 
of the polygon-loop $P_{1}$ at time $\tau$ to another polygon-loop $P_{2}$ 
at the time $\tau +1$.
The functional $k(P)$ is the total curvature of the polygon-loop $P$ which is 
equal to the number of corners of the polygon ( the vertices with 
self-intersection are not counted) and $l(P)$ is the length of $P$ 
which is equal to the number of its links.  The  length functional 
$l(P_{1} \bigtriangleup  P_{2})$ is defined as  
(see formula (12) of \cite{sav2} )
\be
l(P_{1} \bigtriangleup  P_{2}) = l(P_{1}) + l(P_{2}) - 
2 l(P_{1} \cap  P_{2})  \label{dist},
\qee
where the polygon-loop~ $P_{1} \bigtriangleup  P_{2}~ \equiv~ P_{1} \cup  
P_{2} ~\backslash~ P_{1} \cap  P_{2} $~ is a union of links 
$P_{1} \cup P_{2}$ without common links $P_{1} \cap  P_{2}$. The operation
$\bigtriangleup$ maps two polygon-loops $P_{1}$ and $P_{2}$
into a  polygon-loop $P= P_{1} \bigtriangleup  P_{2}$. Note that the  
operations $\cup$ and $\cap$ do not have this property. 
These operations acting on a polygon-loops can produce link 
configurations 
which do not belong to $\Pi$.\footnote{The symmetric difference 
of sets $P_{1} \bigtriangleup  P_{2}$ is an important concept in 
functional analysis \cite{kolmogorov}.} The 
length functional $l(P_{1} \bigtriangleup  P_{2})$  defines a 
distance between two  polygon-loops $P_{1}$ and $P_{2}$. 
It is natural
that the transition amplitude $K_{\beta}(P_{1},P_{2})$ (\ref{tran}) depends 
only on 
the functional which mesures the distance between the 
two subsequent polygon-loop
configurations, because transition amplitude decreases when the distance 
(\ref{dist})
between two configurations increases. This can be seen also from the 
inequality
\be
 l(P_{1} \bigtriangleup  P_{2})~ \geq ~\vert l(P_{1}) - 
l(P_{2}) \vert \label{ineq},
\qee
therefore
\be
K^{gonihedric}_{\beta}(P_{1},P_{2}) \leq  
exp\{ -2\beta ~\vert l(P_{1}) - l(P_{2}) \vert ~ \}   \label{ineqker}.
\qee
Algebraically one can construct
many functionals of that kind, but what is important here is that this distance 
functional appears naturally from the geometrical 
action (\ref{part}) of the original theory. The expression for the transition 
amplitude $K_{\beta}(P_{1},P_{2})$ in the polygon-loop space $\Pi$ is very close 
in its form with the transfer matrix for the random walks 
\be
 K_{\beta}(X,Y) =   exp\{ -\beta \vert X-Y \vert \}  ,
\qee
which depends only on  the distance  between initial 
and final position
of the point patricle. Statistical mechanics of paths with curvature-dependent 
action is also well known  \cite{ambjorn}.
The aim of this work is to study spectral properties of 
the transfer matrix $K_{\beta}(P_{1},P_{2})$ and critical behaviour of the 
statistical system (\ref{trace}).

\section{Free energy and correlation functions}
The eigenvalues of the transfer matrix 
$K_{\beta}(P_{1},P_{2})$ define all 
statistical properties of the system and can be found as a solution of the 
following integral equation in the loop space $\Pi$\footnote{We shall use the 
word "loop" for the "polygon-loop".}

\be
\sum_{\{P_{2}\} }K_{\beta}(P_{1},P_{2})~\Psi(P_{2})=
\Lambda(\beta)~\Psi(P_{1})  \label{inte},
\qee
where $\Psi(P)$ is a function on loop space. The Hilbert space of 
compex functions $\Psi(P)$ on $\Pi$ will be denoted as
$H=L^{2}(\Pi)$.

The eigenvalues define the partition function (\ref{trace})

\be
Z(\beta)= \Lambda^{N}_{0}+...
+\Lambda^{N}_{\gamma} 
 \label{limit},
\qee
and in the thermodynamical limit the free energy is equal to 
\be
-\beta~ f(\beta) = \lim_{N \rightarrow \infty} 
\frac{1}{N^3}~ln~ Z(\beta)    .
\qee
The correlation lengths are defined by the ratios of eigenvalues 
$\Lambda_{i}(\beta)/\Lambda_{0}(\beta)$
\be
\xi_{i}(\beta) = \frac{1}{-ln \frac{\Lambda_{i}(\beta)}
{\Lambda_{0}(\beta)}} \label{corr},
\qee
and grow if the eigenvalues $\Lambda_{i}(\beta)$ approach the eigenvalue 
$\Lambda_{0}(\beta)$ at some critical temperature $\beta_{c}$. By  
the Frobenius-Perron theorem $\Lambda_{0}(\beta)$ is simple and  we have 
\be
\Lambda_{0}(\beta) > \Lambda_{1}(\beta) \geq \Lambda_{2}(\beta) \geq ...
\qee
Finite time propagation amplitude of an initial loop $P_{i}$ to 
a final loop $P_{f}$ for the time interval  $t = M/ \beta$ can be 
defined as 

\be
K(P_{i},P_{f}) = \Lambda^{-M}_{0} 
\sum_{\{P_{1},P_{2},...,P_{M-1}\}}~ K_{\beta}(P_{i},P_{1})\cdots 
K_{\beta}(P_{M-1},P_{f}), \label{prop}
\qee
where we have introduced natural normalization to the biggest eigenvalue 
$\Lambda_{0}$ and $M \leq N$. The trace of the operator $K(P_{i},P_{f})$ is 
equal to 
\be
TrK  = (1 + (\Lambda_{1}/ \Lambda_{0})^{M} +\cdots + 
(\Lambda_{\gamma}/ \Lambda_{0})^{M})
\qee
and depends on the ratio $\Lambda_{i}/ \Lambda_{0}$.

\section{Connection between 3D gonihedric system and 2D-Ising model}
We consider below a transfer matrix which is less 
complicated than the original matrix (\ref{tran}) and depends  
only on the distance functional $l(P_{1} \bigtriangleup  P_{2})$ \footnote{It is 
convinient to substruct the vacuum energy $2 N^2$ in the exponent.}
\be
K_{\beta}(P_{1},P_{2}) = 
exp\{ -2\beta ~(~l(P_{1} \bigtriangleup  P_{2}) -  N^2~) ~ \}.  \label{simp}
\qee
The largest eigenvalue appearing in the equation 
\be
\sum_{\{P_{2}\}}~exp\{ -2\beta  
l(P_{1} \bigtriangleup  P_{2}) + 2\beta N^2 ~ \}\cdot  \Psi(P_{2})=
\Lambda(\beta)~\Psi(P_{1}) \label{eige}
\qee
can be found because the corresponding eigenfunction is a constant function
$\Psi_{0}(P)=1$ (see Appendix). Therefore 
\be
\Lambda_{0} =  \sum_{\{P_{2}\}}~exp\{ -2\beta  
l(P_{1} \bigtriangleup  P_{2}) + 2\beta N^2 ~ \}.  \label{larg}
\qee
The sum (\ref{larg}) does not depend on $P_{1}$. To prove this we note that the 
loop $P = P_{1} \bigtriangleup  P_{2}$ runs over all loops in $\Pi$ as $P_{2}$
runs over $\Pi$, i.e. the mapping $P_{1} \rightarrow P_{1} \bigtriangleup  P_{2}$
is one to one for any $P_{2}$. This change of the variable proves that
\be
\Lambda_{0} =  \sum_{\{P\}}~exp\{ -2\beta  
l(P) + 2\beta N^2 ~ \},  \label{2dis}
\qee
so $\Lambda_{0}$ is the partition function of the 2D-Ising ferromagnet. 
Indeed \cite{kramers,onsager,kac}, 
\be
Z_{2D-Ising} =  \sum_{\{P\}}~exp\{ -2\beta  
l(P) + 2\beta N^2 ~ \}= e^{-\beta f(\beta)\cdot N^2},  \label{ising}
\qee
where $f(\beta)$ is the free energy of the 2D-Ising model. Therefore 
\be
\Lambda_{0} ~=~ Z_{2D-Ising}~=~\lambda^{N}_{0} +...+ \lambda^{N}_{2^N},
 \label{eign0}
\qee
where $\lambda_{i}$ are the eigenvalues of the transfer matrix of 2D-Ising 
model \cite{onsager,kac,baxter}.
{\it Thus the largest eigenvalue of the 3D-system (\ref{simp}) is  
equal to the partition function of the 2D-Ising ferromagnet}. 

The free energy of the 2D-Ising ferromagnet in the thermodynamical limit is 
given by \cite{onsager,kac,baxter}
\be
-\beta f(\beta) =  \frac{1}{2}\int^{2\pi}_{0} 
\frac{d\xi d\eta}{(2\pi)^2}~ln[~(1+w^{4})^{2} -2 (w^{2} -w^{6}) 
(\cos \xi + \cos \eta )~] 
,  \label{2disf}
\qee
where $w =e^{-\beta}$. 
Therefore the free energy of the three-dimensional system 
which is defined by the transfer matrix (\ref{simp}) is given by $f(\beta)$
and coincides with the one of 2D-Ising ferromagnet. 

From this result we can deduce that the critical temperature of the 
three-dimensional system (\ref{simp}) is equal to the one for the 
2D-Ising ferromagnet $2 \beta_{c} = ln(\sqrt{2} - 1)$, that the specific 
heat exponent $\alpha  =0$ and from the hyperscaling law 
$\nu d = 2 -\alpha$ that $\nu = 2/3$.

We recall \cite{sav2} that in the alternative approximation 
when the intersection term 
$2k(P_{1}\cap P_{2})$ is ignored in the original transfer matrix (\ref{tran})  
\be
K_{0\beta}~(P_{1},P_{2})=
exp\{ -\beta  ~[k(P_{1} \bigtriangleup  P_{2}) + 
2 l(P_{1} \bigtriangleup  P_{2}) - 2N^2 ]~  \}~~  \label{appro}
\qee
the free energy can also be computed \cite{sav2}
$$
-\beta f_{0}(\beta) =  \frac{1}{2}\int^{2\pi}_{0} 
\frac{d\xi d\eta}{(2\pi)^2}
ln[~(1+w^{4})^{2} - 4w^{8}\omega^{2}(1-\omega^{2}) + 
$$
\be
+4w^{4}(1-\omega^{2})\cos \xi \cos \eta - 
2(w^{2} - 2  w^{6} \omega^{2} + w^{6})(\cos \xi  + \cos \eta )~], 
  \label{oldr}
\qee
where $\omega^{2}=w$ is the contribution from the curvature term $k(P)$
and exhibits the same critical behaviour 
as the 2D-Ising ferromagnet. Thus the original gonihedric system (\ref{trace}),
(\ref{tran}) is bounded by two close statistical systems  
\be
-\beta f_{0}(\beta) \leq   -\beta f_{gonihedric}(\beta)     
\leq  -\beta f(\beta),
\qee
because
\be
 K_{0\beta}~(P_{1},P_{2})~\leq~ K^{gonihedric}_{\beta}~(P_{1},
P_{2})~\leq~ K_{\beta}~(P_{1},P_{2})  \label{appro1}
\qee
This confirms the conjecture \cite{sav2} that 3D gonihedric system should 
have 
statistical properties close to the ones of 2D-Ising ferromagnet. Earlier 
numerical simulations \cite{bathas,johnson,cut} support this dimensional 
"reduction". It is 
also consistent with the analytical estimate of the enthropy factor of the 
random surfaces on a cubic lattice \cite{pie,wegpie}. 

Layer-to-layer transfer matrices for three-dimensional statistical systems, 
whose elements are  the product of all Boltzmann weight functions of cubes
between two adjacent layers have been considered in the literature 
\cite{polyakov,fradkin,zamolodchikov,bazhanov}. 
Using Yang-Baxter and Tetrahedron equations 
one can compute the spectrum of the transfer matrix in a number of interesting 
cases \cite{zamolodchikov,bazhanov}. In the given case the transfer matrix 
(\ref{tran}) has geometrical interpretation which helps to compute the spectrum.
\vspace{0.3cm}

We have to remark that if the loops $\{P\}$
are not restricted to be closed  the system is essentially simplified.
Indeed in the model
where all subsets of links from $T^2$ are allowed as configurations of 
the system and the transfer matrix is defined by the same formula 
the model becomes trivial, the transfer matrix is the $2N^2$th tensor 
product of 1D-Ising model transfer matrix
\be
K_{\beta} = \left( \begin{array}{c}~~~1~~~~~~~~~e^{-2\beta}
\\~~e^{-2\beta}~~~~~~1~~~~\end{array} \right).
\qee
Using this observation one can derive inequality
$$
\Lambda_{1} ~\leq ~\frac{(1+e^{-2\beta})^{2N^2 -1}(1-e^{-2\beta})}{Z_{2D-Ising}}.
$$

\section{Loop space and eigenfunctions}
To proceed it is convenient to introduce some notation. 
An invariant product in $\Pi$ can be defined as  

\be
<P_{1}\vert P_{2}> ~~= ~~l(P_{1} \bigtriangleup  
\bar{P_{2}})~ - ~l(P_{1} \bigtriangleup  P_{2})  \label{inva}
\qee
where $\bar{P} = T^{2}~\bigtriangleup P~=~ T^{2}~\backslash ~P$ and $T^{2}$ 
is the loop which contains all links of the toroidal lattice $T^{2}$ 
($~l(T^{2}) = 2N^2$~). The invariant product (\ref{inva}) is also equal to 
\be
<P_{1}\vert P_{2}> ~~= ~~2N^{2} ~-  2~l(P_{1} \bigtriangleup  P_{2}) 
\label{invaeq}
\qee
and is  odd with respect to $\bar{P}$-the complement 
operation $<P_{1}\vert P_{2}>~=~-<P_{1}\vert \bar{P_{2}}>$.
In particular the energy functional of the system is equal to 
\be
E_{P}~=~<0~\vert~ P>~=~2N^2 - 2 l(P)         \label{rho3}
\qee
The product (\ref{inva}) is invariant under the simultaneous $rotations$
 of the loops $P_{1}$ and $ P_{2}$, defined as

\be
P~~ \rightarrow~~ P \bigtriangleup  \delta,
\label{trans}
\qee
where $\delta$ is an arbitrary loop
\be
<P_{1}\bigtriangleup  \delta ~\vert ~P_{2} \bigtriangleup  
\delta> ~~= ~~<P_{1}\vert P_{2}> . \label{traninva}
\qee
This group of transformations in the loop space $\Pi$ is Abelian because for its 
repesentations on H
\be
R_{\delta}  ~\Psi(P) ~=~ \Psi(P\bigtriangleup \delta) ,\label{repre}
\qee
we have 
\be
[R_{\delta_{1}},~R_{\delta_{2}}]~=~0~~~~~~~~ R^{2}_{\delta}=1.
\label{repres}
\qee
The product $<P_{1}\vert P_{2}>$ is invariant also under $translations$. The
group of translations is defined as a rigit translation of the loop P in x and y 
directions on $T^2$ by same units of lattice spacing $a$
\be
P~~ \rightarrow~~ P  + a_{ x}e_{x} + a_{y}e_{y}. \label{ro}
\qee
Together these two groups 
form a Nonabelian group which acts on the loop space $\Pi$.

With this notation the transfer matrix (\ref{simp}) takes the form
\be
K_{\beta}(P_{1},P_{2}) = 
exp\{ ~\beta ~<P_{1}\vert P_{2}>~ \}.  \label{kern}
\qee
and the integral equation (\ref{eige}) the form 
\be
\sum_{\{P_{2}\}}~e^{\beta  ~<P_{1}\vert P_{2}>} \cdot  \Psi(P_{2})=
\Lambda(\beta)~\Psi(P_{1}). \label{eigngen}
\qee
Note that (\ref{eigngen}) is invariant under rotations (\ref{trans}). 
This suggests that we search for eigenfunctions of the operator 
(\ref{kern})  in the form
of invariant polynomials in $H=L^{2}(\Pi)$. These polynomials can be constructed
in the form of powers of the invariant $<P \vert Q>$ as follows:
\begin{eqnarray}
\Psi^{(0)}_{Q}(P) = 1,~~~~~~~~~~~~~~~~~~~~~~~~~~~~~~~~~ \nonumber    \\
\Psi^{(1)}_{Q}(P) = <P \vert Q>, ~~~~~~~~~~~~~~~~~~~~~~~\nonumber    \\
\Psi^{(2)}_{Q}(P) = <P \vert Q>^{2}~ - ~\rho_{2},~~~~~~~~~~~~~ \nonumber    \\
\Psi^{(3)}_{Q}(P) = <P \vert Q>^{3}~ - ~\rho_{3}~<P \vert Q>, \nonumber    \\
..........................................................\nonumber    \\
\Psi^{(n)}_{Q}(P) =
 <P \vert Q>^{n} ~-~\rho_{n}~<P \vert Q>^{n-2}-...\label{bases}
\end{eqnarray}
where the coefficients
\vspace{.5cm}

\be
\rho_{2} =\frac{ \sum_{\{P\}} ~<0~\vert~ P>^{2} }
{\sum_{\{P\}}1},~~~~~\rho_{3} =\frac{\sum_{\{P\}}~<0~\vert~ P>^{4}}
{\sum_{\{P\}}~<0~\vert~ P>^{2}},  \label{rho2}
\qee
and so on are chosen so that $\Psi^{(i)}_{Q}$  is orthogonal to 
$\Psi^{(j)}_{Q}$ if 
$i\neq j$. The set of functions $\{\Psi^{(n)}_{Q}(P)\}$ in $H$ form 
an invariant 
subset $H_{n}$  of the level $n$. The index $Q$ numerates functions
inside the level. We introduce general Legandre loop polynomials as 
\be
L_{n}(x) = x^n -   \rho_{n}x^{n-2}-...... \label{rhon}
\qee
so $L_{n}(<P \vert Q>)~=~\Psi^{(n)}_{Q}(P)$.
This set of functions in $H$  is 
appropriate for solving  the integral equation (\ref{eigngen}) and in its 
form is very similar with the ones which are used for the random paths with 
curvature-dependent action \cite{ambjorn}. 

Let us begin by
proving that $\Psi^{(1)}_{Q}(P)$ is the eigenfunction of (\ref{eigngen}). Because
(see the next section)
\be
\sum_{\{P\}}~< Q_{1}\vert P >^{n} \cdot  
< P \vert Q_{2}> ~=~\mu_{n1}~<Q_{1}\vert Q_{2}>  \label{norm}
\qee
where
\be
\mu_{n1} = \frac{1}{2N^2}~\sum_{\{P\}}~<0~\vert~P>^{n+1},~~~~~\mu_{2k~1}=0 
\label{normcoef}
\qee
we have  
\be
\sum_{\{P_{2}\}}~e^{\beta  ~<P_{1}\vert P_{2}>} \cdot  <P_{2} \vert Q> =
\Lambda_{1}(\beta)~ <P_{1} \vert Q> \label{eign1} 
\qee
where 
\be
\Lambda_{1}  = \sum^{\infty}_{n=0} \frac{\beta^{n}}{n!}\mu_{n1} =
\frac{1}{2N^2}~\sum_{\{P\}}~exp\{~ \beta~<0~\vert~P>~\}~<0~\vert~P>  \label{eign11}
\qee
and thus
\be
\Lambda_{1}  = \frac{1}{2N^2}~\frac{\partial}{\partial \beta} 
\Lambda_{0} .\label{eign111}
\qee
For the ratio $\Lambda_{1}/\Lambda_{0}$ one can get
\be
-(\Lambda_{1}/\Lambda_{0}) = \frac{-1}
{2N^2}~\frac{\partial}{\partial \beta}~ln~\Lambda_{0}~= ~u(\beta) 
\label{inenerg}
\qee
which is the internal energy of the 2D-Ising system. {\it Thus the second
eigenvalue of the three-dimensional system coincides with internal energy of 
the two-dimensional Ising system.}

Using equation (\ref{norm}) one can also address the question of degeneracy of
the eigenvalues. The scalar product of the loop functions can be defined as
\be
\Psi_{Q_{1}}\cdot \Psi_{Q_{2}} = 
\sum_{\{P\}} \Psi_{Q_{1}}(P)~ \Psi_{Q_{2}}(P)
\label{scalar}
\qee
then for the zero and first level functions (\ref{bases}) we have 
$$
\Psi^{(0)}_{Q_{1}}\cdot \Psi^{(1)}_{Q_{2}} =0,
$$
\be
\Psi^{(1)}_{Q_{1}}\cdot \Psi^{(1)}_{Q_{2}}~ =~ \mu_{11} <Q_{1}\vert Q_{2}>.
\label{orthog}
\qee
\vspace{.2cm}
The rank of the matrix $G_{Q_{1}Q_{2}}=<Q_{1}\vert Q_{2}> $ defines the number
$\epsilon_{1}$ of linearly independent functions on the first level $H_{1}$ 
and thus the degeneracy of the eigenvalue $\Lambda_{1}$ 
\footnote{The degeneracy  of the zero level is one $\epsilon_{0}=1$.}
The number $\epsilon_{1}$
is bigger or equal to the number of functions $\Psi^{(1)}_{Q_{i}}$ which 
are orthogonal to each other. The orthogonality condition follows from 
(\ref{orthog})
$$
<Q_{i}\vert Q_{j}>~=~0~~~~~~~~i,j=1,2,....,2N^2
$$
The last equation can be rewritten also in the form
$$
l(Q_{i} \bigtriangleup  Q_{j}) = N^2
$$
and its solutions provide $2N^2$ linearly independent functions on the first level.
The solutions will be presented in a separate place.

In the following we will search for approximate solutions since the sums 
$$
\sum_{\{P\}}~< Q_{1}\vert P >^{n} \cdot  < P \vert Q_{2}>^{k},~~~~k \geq 2
$$
depend not only on invariant $<Q_{1}\vert Q_{2}>$, but also on the shape of 
$Q_{1}$ and $Q_{2}$. However the sum depends only weakly on the shape of $Q's$
and the leading behaviour involves only the length. In this approximation 
we have (see also the next section)
\be
\sum_{\{P\}}~< Q_{1}\vert P >^{n} \cdot  
< P \vert Q_{2}>^{2} ~=~\mu_{n2}<Q_{1}\vert Q_{2}>^{2}~ + ~\mu_{n0} 
\label{norm0}
\qee
where
$$
\mu_{n2} ~=~\frac{1}{(2 N^2)^2 - \rho_{2} } \sum_{\{P\}}
(< 0 \vert P >^{n+2}- \rho_{2}< 0 \vert P >^{n})
$$
and 
$$
\mu_{n0} ~=~-\frac{\rho_{2}}{(2 N^2)^2 - \rho_{2}}
\sum_{\{P\}}(< 0 \vert P >^{n+2}- (2 N^2)^{2}< 0 \vert P >^{n})
$$
so that the next eigenvalue can be found in the same way
\be
\Lambda_{2}  = \frac{1}{(2 N^2)^2 - \rho_{2} } 
\sum_{\{P\}}~exp\{~ \beta~ <0~\vert~P>  ~\}~[<0~\vert~P> ^2  - \rho_{2}] ,    \label{eign2}
\qee
and for the third one we have
\be
\Lambda_{3}  = \frac{1}{(2 N^2)^3 - \rho_{3} (2N^2)} 
\sum_{\{P\}}~exp\{~ \beta~(<0~\vert~P> ) ~\}~[<0~\vert~P> ^3  - 
\rho_{3}<0~\vert~P> ] , \label{eign3}
\qee
and so on. All eigenvalues can be expressed through the generalysed 
Legandre polynomials (\ref{rhon})
\be
\Lambda_{n}  = \frac{1}{L_{n}(2N^2)} 
\sum_{\{P\}}~exp\{~ \beta~<0~\vert~P> ~\}~L_{n}(<0~\vert~P> ) , \label{eignn}
\qee
or as corresponding derivative of the Ising partition function
\be
\Lambda_{n}(\beta)  = \frac{1}{L_{n}(2N^2)}~ L_{n}(\frac{\partial}
{\partial \beta})~\Lambda_{0}(\beta) , \label{eignnn}
\qee
This completes the computation of the eigenfunctions and eigenvalues 
of the integral equation (\ref{eige}), (\ref{eigngen}).
\vspace{1cm}

\section{Loop correlation functions}
The equality (\ref{norm}) can be proven by using the fact that
the l.h.s. of (\ref{norm}) is equal to 
$$
\sum_{\{P\}}~< 0 \vert P >^{n} \cdot  
< P \vert Q> ~=~\sum_{\{P\}}~< 0 \vert P >^{n+1} -2l(Q)
\sum_{\{P\}}~< 0 \vert P >^{n} + 
$$
\be
+4  \sum_{\{P\}}~< 0 \vert P >^{n}\cdot~ l(P \cap Q).  \label{equal}
\qee
The last term in (\ref{equal}) is a linear function of $l(Q)$. 
If $Q= Q_{1}\cup Q_{2}$ and $Q_{1}\cap Q_{2} = 0$ then
$l(P \cap Q) = l(P \cap Q_{1}) + l(P \cap Q_{2})$ and we have
\be
\sum_{\{P\}}~< 0 \vert P >^{n}\cdot~ l(P \cap Q)~=~\eta_{n1}~l(Q).  
\label{equal1}
\qee
The normalization constant can be computed at the point $Q=T^2$
\be
\eta_{n1}~=~\frac{1}{2N^2}\sum_{\{P\}}~< 0 \vert P >^{n} l(P).  
\label{equal11}
\qee
The linearity of the last term in (\ref{equal}) with respect to $l(Q)$
follows also from the 
decomposition of the functional $l(P \cap Q)$ into the sum over the 
links of $Q$ 
$$
l(P \cap Q) = \sum_{i \in Q} \chi_{i}(P)
$$
where $\chi_{i}(P)=1$ if $i \in P$ and zero otherwise.
By homogenuity 
$$
\sum_{\{P\}}~< 0 \vert P >^{n}\cdot~ \chi_{i}(P) 
$$
is independent of $i$. Hence the summation over $i \in Q$ gives the expected 
result (\ref{equal1}).

To proceed we have to compute the sum 
\be
\sum_{\{P\}}~< Q_{1}\vert P >^{n} \cdot  
< P \vert Q_{2}>^{2},  \label{norm1}
\qee
which by the transformation similar to (\ref{equal})
can be expressed in terms of the {\it two-set correlation function} 
\be
\sum_{\{P\}}~< 0 \vert P >^{n}~ \cdot ~ l(P \cap Q_{1}) \cdot  
l(P \cap Q_{2}). \label{norm2}
\qee
The last correlation function is a sum over $i \in Q_{1}$ and $j \in Q_{2}$
of the two-link correlation function 
\be
\sum_{\{P\}}~< 0 \vert P >^{n}\cdot~ \chi_{i}(P)~\chi_{j}(P)\label{norm3}.
\qee
In the approximation when the two-link correlation function (\ref{norm3})
factorises, then the correlation function (\ref{norm2}) is equal to
$$
\eta_{n2} ~l(Q_{1}) l(Q_{2})~+~ \eta_{n0}~l(Q_{1} \cap Q_{2})
$$
and we recover the expression (\ref{norm0}).

For the high powers we have to compute a many-set correlation function 
\be
\sum_{\{P\}}~< 0 \vert P >^{n} \cdot \prod_{a} l(P \cap Q_{a}) 
~=~\eta_{na} ~\prod_{a} l(Q_{a})~+~\cdots   \label{norm4}
\qee
which can be expressed as a function of lengths in the approximation 
when many-link correlation function 
\be
\sum_{\{P\}}~< 0 \vert P >^{n} \cdot~\chi_{i}(P)~\chi_{j}(P)~\chi_{k}(P)\cdots,
\label{norm5}
\qee
factorises. Because the link variable $\chi_{i}(P)$ is a quadratic function 
of Ising spins it follows that many-link correlation function 
(\ref{norm5}) is nothing else than many-spin correlation function of the 
2D-Ising model.

\section{ Acknowledgement} One of the authors (T.J.) is indebted to the 
National Research Center Demokritos for hospitality.
This work was supported in part by the EEC Grant no. ERBFMBICT972402.

\section{Appendix }
It is elementary to verify that 
the function $l(P_{1} \bigtriangleup  P_{2})$ (\ref{dist}) on 
$\Pi \times \Pi$ is a distance functional $\rho(P_{1}, P_{2}) \equiv 
l(P_{1} \bigtriangleup  P_{2})$ and that if $P_{1}$, $P_{2}$, 
$P \in \Pi$ are such that 
$l(P_{1} \bigtriangleup  P_{2}) = l(P_{1} \bigtriangleup  P) + 
l(P \bigtriangleup  P_{2})$, then $P_{1} \cap  P_{2} \subseteq P \subseteq 
P_{1} \cup  P_{2}$. Indeed we have 

$$
l(P)  = l(P \cap  P_{1})  -2 l(P_{1} \cap  P_{2}) + l(P_{2} \cap  P),
$$
hence
$
l(P \backslash P_{1} \cup  P_{2})  = l(P \cap  P_{1} \cap P_{2})  -
l(P_{1} \cap  P_{2}) 
$
from which we conclude that $P \subseteq P_{1} \cup  P_{2}$ and 
$P_{1} \cap  P_{2} \subseteq P $.
\vspace{.3cm}

We complete our arguments by showing that all eigenvalues of the normalized 
transfer matrix $\Lambda^{-1}_{0}K_{\beta}(P_{1},P_{2})$ are smaller than or 
equal to 1. In order to prove this statement let $\chi_{P}$ be a "delta function" 
in $\Pi$, i.e. $\chi_{P}(Q) =1 ~if~ P=Q$ and $\chi_{P}(Q) =0$~otherwise. The 
set of functions $\{ \chi_{P} \}$ form a basis in $H=L^2 (\Pi)$. Note that
for the propagator (\ref{prop})
$$
K(P_{i},P_{f}) = <\chi_{P_{i}}~\vert 
\Lambda^{-M}_{0} K^{M}_{\beta}\vert ~\chi_{P_{f}}>
$$
we have (\ref{prop})
$$
K(P_{i},P_{f}) = \Lambda^{-M}_{0} 
\sum_{\{P_{1},P_{2},...,P_{M-1}\}}~ K_{\beta}(P_{i},P_{1})
K_{\beta}(P_{1},P_{2})\cdots K_{\beta}(P_{M-1},P_{f}) \leq
$$
$$
\leq~\Lambda^{-M}_{0} 
\sum_{\{P_{1},P_{2},...,P_{M-1}\}}~ K_{\beta}(P_{i},P_{1})
K_{\beta}(P_{2},P_{3})\cdots K_{\beta}(P_{M-1},P_{f}) .
$$
We can sum successively over $P_{i}'s$ to obtain
$$
K(P_{i},P_{f})~ \leq ~\Lambda^{-1}_{0}
$$
for any $P_{i},P_{f}$ provided $M>1$.  If $K(P_{i},P_{f})$ had an eigenvalue
$\Lambda \geq 1$ then $K(P_{i},P_{f})$ would grow as $\Lambda^{M}$ as 
$M \rightarrow \infty$ for same $P_{f}$. Since $K_{\beta}(P_{1},P_{2}) >0$ it 
follows from the Frobenius-Perron theorem that 1 is a simple eigenvalue.

It follows that for $M > 1$,~~$
K(P, P) \leq \Lambda^{-1}_{0},
$
thus
$$
\frac{ln~ \Lambda_{0}(\beta)}{M}  \leq m(\beta),
$$
where the mass $ m(\beta)$ is defined as 
$$
 K(P, P) \approx exp \{~-M~ m(\beta)~\}
$$
for sufficiently large $M$.
From the other side if in the sum which defines $K(P, P)$ we restrict all 
intermediate loops to be $P$ then~$
\Lambda^{-M}_{0}(\beta) ~ \leq ~  K(P, P)~$
and thus
$$
\frac{ln~ \Lambda_{0}(\beta)}{M} ~ \leq~  m(\beta) ~\leq ~ln ~\Lambda_{0}(\beta).
$$

The operator $K(P, P)$ is bounded from below also by positive number 
$1/\gamma $. Indeed, because
$$
K(P, P) = <\chi_{P}~\vert \Lambda^{-M}_{0}K^{M}_{\beta}\vert ~\chi_{P}> = 
\sum_{n=0} <\chi_{P}~\vert \Lambda^{-M}_{0}K^{M}_{\beta}\vert ~\Psi^{(n)}>
<\Psi^{(n)} \vert \chi_{P}> =
$$
$$
=\sum_{n=0} (\frac{\Lambda_{n}}{\Lambda_{0}})^{M} 
\vert <\chi_{P}~\vert ~\Psi^{(n)}>  \vert^2 ~\geq~ \vert 
<\chi_{P}~\vert ~\Psi^{(0)}>  \vert^2 .
$$ 
Since for the normalized constant function on $\Pi$ 
$$
<\chi_{P}~\vert ~\Psi^{(0)}> = 1/ \sqrt{\gamma} ,
$$
where $\gamma$ is total number of loops on a toroidal lattice $T^{2}$,
thus
$$
1/\gamma ~\leq ~K(P, P).
$$
For the mass $ m(\beta)$ one can get 
$$
m(\beta) ~\leq ~\frac{ln \gamma }{M}.
$$

\vfill
\end{document}